\documentclass[prl,reprint]{revtex4-2}
\usepackage{ae}
\usepackage[pdftex]{graphicx}
\usepackage{amsmath}
\usepackage{amssymb}
\usepackage{epsfig, float,color}
\usepackage{hyperref}
\usepackage{color}
\usepackage[dvipsnames]{xcolor}

\begin{document}
\date{\today }

\title{Shockwaves and turbulence across social media}
\author{Pedro D. Manrique, Frank Huo, Sara El Oud, Minzhang Zheng, Lucia Illari, Neil F. Johnson}

\affiliation{Physics Department, George Washington University, Washington, DC 20052, U.S.A.}

\begin{abstract}
 Online communities featuring `anti-X' hate and extremism, somehow thrive online despite moderator pressure. We present a first-principles theory of their dynamics, which accounts for the fact that the online population comprises diverse individuals and evolves in time. The resulting equation represents a novel generalization of nonlinear fluid physics and explains the observed behavior across scales. Its shockwave-like solutions explain how, why and when such activity rises from `out-of-nowhere', and show how it can be delayed, re-shaped and even prevented by adjusting the online collective chemistry. This theory and findings should also be applicable to anti-X activity in next-generation ecosystems featuring blockchain platforms and Metaverses.
\end{abstract}

\maketitle

\noindent Society is struggling with online anti-X hate and extremism, where `X' can nowadays be any topic, e.g. religion, race,  ethnicity \cite{Gill,hategroups,UN, ethiopia,myanmar}. Recent research has confirmed that in-built online communities play a key role in developing support for a topic at scale \cite{SAHD} and anti-X sentiment is no different \cite{Gill,hategroups,UN,ethiopia,myanmar}. These in-built communities are referred to differently on different platforms, e.g. Group on VKontakte and on Gab, Page on Facebook, Channel on Telegram, and are unrelated to community-detection in networks. Each in-built community is a self-organized aggregate of anywhere from a few to a few million users. 
 
 Such anti-X communities can grow quickly from out of nowhere because of interested individuals or other communities joining (fusing) with them (Fig. 1(a), empirical fusion) \cite{SAHD,Forsyth,PsycholFear2005,PsycholFear2011,PsycholBook2021}. Having content that violates platforms' Terms and Conditions means that they can also suddenly get shut down when discovered by moderators (Fig. 1(b), empirical total fission). Therefore in contrast to communities such as pizza fans, there is a clear benefit for such anti-X communities to grow in a bottom-up way in order to remain under moderators' radar. Figure 2(a)(b) illustrates the sea of erratic shark-fin-shaped waves that emerges: each shows an anti-X community's size of membership as it suddenly appears and grows through fusion and may then suddenly disappear via total fission. Some social scientists \cite{PolitTurbul2016} are suggesting that such volatility is `online turbulence' which could -- if proven true -- open up an important new field for physics and also help bridge the current gap between computational approaches to online (mis)behavior and in-depth case-studies  \cite{litreview}.
 
 Unfortunately such physics does not yet exist, i.e. there is no first-principles theory that accounts for populations of objects (e.g. anti-X individuals) that (1) have their own internal character that may evolve over time, {\em and} (2) interact in a distance-independent way as allowed by the Internet, {\em and} (3) have a changing total size (e.g. Internet use jumped $13.2\%$ in 2020), {\em and} (4) undergo rapid fusion-fission dynamics as in Fig. 1, Fig. 2 (a)(b).
  
\begin{figure}
\centering
\includegraphics[width=1.0\linewidth]{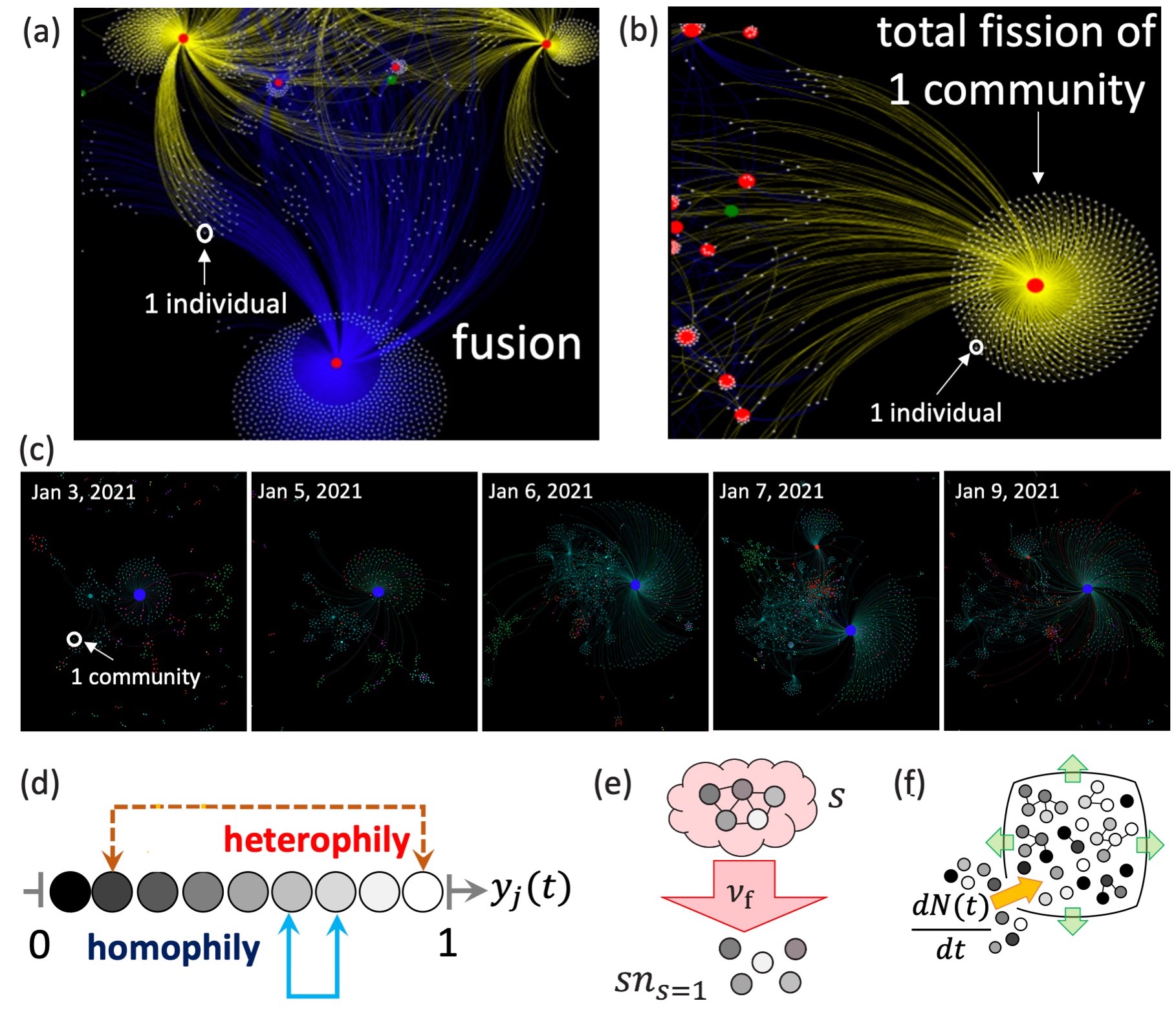}
\caption{\small{Empirically observed (a) fusion and (b) total fission of in-built communities featuring anti-U.S. hate on VKontakte between day $t$ (yellow) and $t+1$ (blue). Red nodes are anti-U.S. communities that later got shut down (total fission); green nodes are those still not yet shut down; yellow links point to individuals (white dots) removed from the anti-U.S. community on day $t+1$; blue links point to individuals added to the anti-U.S. community on day $t+1$. Spatial layout results from (a) and (b) being closeups of a fuller network plotted using ForceAtlas2, meaning that nodes appearing closer together are more interconnected. (b) also shows that very few individuals are simultaneously also members of other communities (SM shows further proof). (c) Empirically observed clustering of anti-government communities across platforms around U.S. Capitol attack. (d)-(f): The theory in this paper incorporates (d) heterogeneous individuals aggregating (i.e. fusion) based on character similarity, (e) total fission with probability $\nu_{\rm f}$, (f) time-varying population size $N(t)$.}}
\label{fig1}
\end{figure}

\begin{figure*}
\centering
\includegraphics[width=1.0\linewidth]{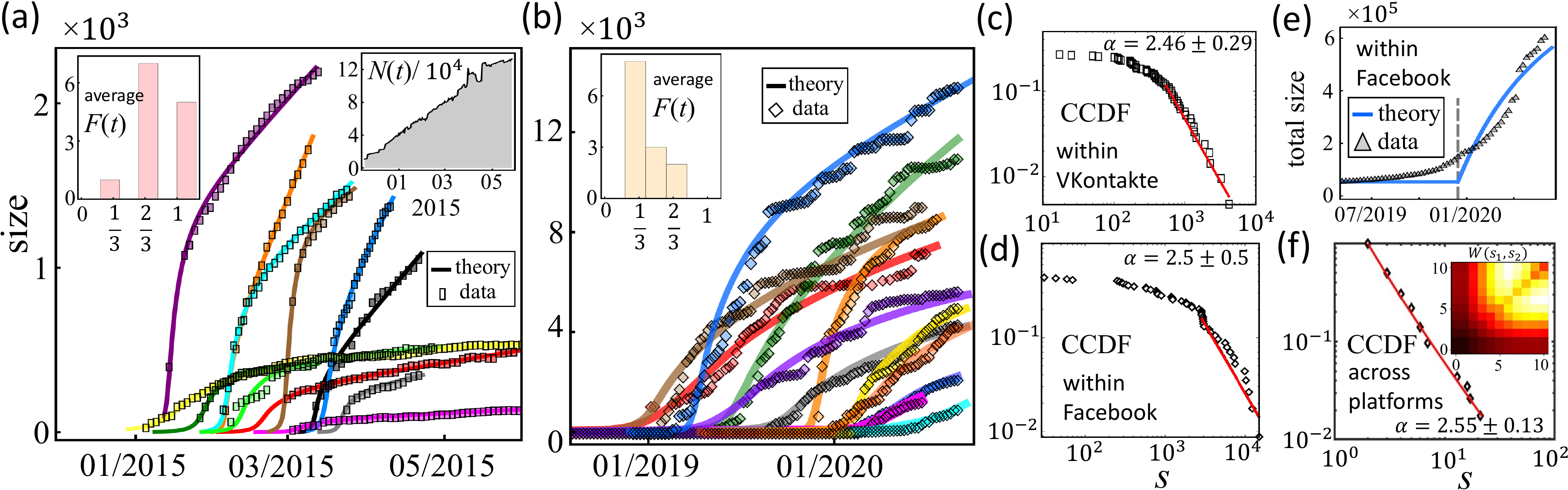}
\caption{\small{Empirical data (symbols) and Eq. 2 theory predictions (lines) for in-built anti-X communities within and across platforms. (a) Size (i.e. number of members) of foreign anti-U.S. (jihadi) communities on  VKontakte. (b) Size of domestic anti-U.S. government (pro-civil war) communities on Facebook. Insets: changing population size; time-averaged $F(t)$ which suggests that (b) reflects a heterophily fusion mechanism more than (a). (c-d) Complementary cumulative distribution (CCDF) of individual community sizes $s$ from (a) and (b). (e) Evolution of total size of all communities from (b). (f) CCDF at a higher scale, i.e. sizes of clusters of interlinked communities. Inset: empirically inferred interaction kernel $W(s_1,s_2)$ obtained from data across all platforms; axes $s_1$ and $s_2$ are sizes of interacting aggregates.}}
\label{fig2}
\end{figure*}

 Here we propose this new physics via a first-principles dynamical theory of anti-X communities within and across social media platforms. The resulting equation (Eq. 2; derivation SM Sec. 2)  provides a novel generalization of nonlinear fluid physics, including shockwaves and turbulence, and extends the physics of aggregation and fragmentation \cite{HH,Gavrilets,Stockmayer43,Gillespie76,Gillespie77,Hendriks83,Manson91,Wattis99,MichelScience01,Lushnikov06,palla2007quantifying,NewmanBook,RednerBook,EZ,PNAS,Centola,Char1,Char2,Char3,prl18,Hartnett16,Gorski20}. Its solutions explain empirically observed patterns within and across social media platforms (Fig. 2), and predict how the rise of anti-X communities can be delayed, re-shaped and even prevented (Fig. 3). Its approximate analytic solutions can also explain the complex multi-community evolution around the U.S. Capitol attack (Fig. 4(b)). Our empirical data is obtained using a published methodology \cite{VKScience16,BoogSciRep21,johnson2019hidden} that we summarize in SM Sec. 1. We are not claiming that all online anti-X activity will always exhibit the patterns in Fig. 2 and Fig. 4(b), or that all non-anti-X activity never does -- however, the important cases shown here do, while SM Sec. 1.1 shows that non-anti-X communities typically do not. Situations in which the observed anti-X activity does not follow these patterns may therefore be indicative of other mechanisms being at play, e.g. top-down coordination or state actor control.

Our theory considers $N(t)$ heterogeneous individuals that are attracted online by such shocking content and hence could aggregate over time depending on their traits. Each aggregate (i.e. in-built community) then totally fragments with some small probability at each timestep (Fig. 1(d)-(f)). Following prior social science and physics studies \cite{HH,Gavrilets,PNAS,Centola,Char1,Char2,Char3,prl18}, each individual $i=1,2,...,N(t)$ can have an arbitrary number of traits, expressed as a vector $\vec{y}_{i}(t)$ where each component (trait value) lies between 0 and 1; but for notational simplicity we only consider one here. At each timestep, two (e.g. randomly) chosen individuals $i$ and $j$ can fuse together  with a probability that depends on the pair's similarity  $|y_{i}(t)-y_{j}(t)|$  (Fig. 1(d)).
 If $i$ and $j$ are already part of an aggregate, their whole aggregates fuse. Hence the mechanism accounts for loners (aggregates of size $s=1$) joining together, or a loner joining a community (aggregate of any size $s>1$), or two communities (aggregates of any size $s_1,s_2>1$) joining together (Fig. 1(a)). We can calculate a mean-field fusion probability $F(t)$ by averaging over the population distribution at time $t$, e.g. for a constant uniform distribution, pairing favoring similarity (homophily) yields $F(t)=2/3$ while dissimilarity (heterophily) yields $F(t)=1/3$ (details in SM Sec. 2.1). In this way, $F(t)$ captures the online collective chemistry.

Master equations for the number $n_{s}(t)$ of aggregates of size $s$, are for $s>1$ and $s=1$ respectively:
{\small{\begin{eqnarray}
{\dot n}_s
&=&{\frac{F(t)}{N(t)^2}}\sum_{s_1+s_2=s} s_1 n_{s_1} s_2 n_{s_2}-{\frac{2 F(t) s n_s}{N(t)^2}\sum^{\infty}_{s_1=1} s_1 n_{s_1}}\nonumber-{\frac{\nu_{\rm f} s n_s}{N(t)}}
\nonumber\\ 
{\dot n}_1&=&-\frac{2 F(t) n_1}{N(t)^2}\sum^{\infty}_{{s_1}=1} {s_1} n_{{s_1}}+\frac{\nu_{\rm f}}{N(t)}\sum^{\infty}_{s_1=2}s_1^2 n_{s_1} +{\dot N}(t) 
\end{eqnarray}}}

\noindent where the first term(s) on each right-hand side are fusion, the next are total fission, and the final ${\dot n}_1(t)$ term is the influx of potential recruits. 
The fusion product-kernel is justified empirically by studies of humans' electronic communications \cite{palla2007quantifying} and by the online anti-X data (see Fig. 2(f) inset and SM Sec. 1.1). We made the  reasonable assumption that  macrolevel quantities $F(t)$ and $N(t)$ vary slowly compared to  microlevel aggregation: the SM shows this is justified by comparing to full microscopic simulations.
Defining $u(x,t)=\sum_{s}sn_{s}(t)e^{-xs}$, Eq. 1 becomes
\begin{eqnarray}
{\dot u}(x,t)&=&-\frac{2F(t)}{N^2(t)}u(x,t){ u'}(x,t)+\frac{2[F(t)+\nu_{\rm f}/2]}{N(t)}{ u'}(x,t)\nonumber\\
&+&e^{-x}[{\dot N}(t)-\frac{\nu_{\rm f}}{N(t)}{ u'}(x,t)|_{x=0} ]
\label{eq:PDEmain}
\end{eqnarray}
where $u'$ is the $x$-derivative. Equation 2 is a novel generalization of nonlinear fluid equations with shockwave solutions. An additional link mechanism, discussed later, would add the diffusive term $u''(x,t)$ typical of turbulence studies in viscous fluids. If platform moderators are ineffective in implementing shutdowns (i.e. $\nu_{\rm f}\rightarrow 0$) and if $N(t)$ and $F(t)$ are effectively constant, Eq. 2 reduces to the well-known case of the inviscid Burgers equation. The solutions $u(x,t)$ yield the anti-X community size $S(t)=N(t)-u(0,t)$ that develops between shutdown events: in this simple limit $S(t)=N\left(1+(t_{sw}/t)W[-(t/t_{sw})\exp{(-t/t_{sw})}]\right)$ with $W[.]$ the Lambert function, akin to the emergence of a giant connected component in a network interpretation, and its onset time $t_{sw}=N(2F)^{-1}$, i.e. anti-X community will suddenly appear at $t_{sw}$ and grow rapidly. More generally, $t_{sw}\approx{\overline{N(t)}}(2{\overline{F(t)}})^{-1}$ involving time-averages (SM Sec. 2.3.3).

Generalizing to multiple traits per individual, allows multiple anti-X communities to emerge: each has its own `flavor' of the anti-X topic (i.e. the members of different communities are concentrated along a different $\{\vec{y}_{i}(t)\}$-axis, see SM Se. 2.7);  each has its own onset time (i.e. they appear asynchronously as observed empirically in Fig. 2(a)(b)); and each has its own growth curve (again as observed empirically in Fig. 2(a)(b)). This explains why typical observers of social media simply see an erratic succession of different anti-X communities rising suddenly and unexpectedly from `out of nowhere'.

Figures 2(a) and (b) show explicitly the generalized shockwave solutions of Eq. 2 in between shutdown events. They reproduce the complex growth curve shapes for both (a) foreign anti-U.S. communities and (b) domestic U.S. anti-government communities (see SM Sec. 1 for data details). The simpler Burgers' equation solutions give far poorer fits. The inferred average $F(t)$ values shown in the insets are closer to heterophily in (b) than in (a) which is consistent with the highly diverse nature of the support reported for (b) \cite{boogJan6news}. Figures 2(c)(d) show that the theory's predicted 2.5-exponent power-law distribution for the community sizes (see SM Sec. 2.4.2.2 for proof) is also consistent with the empirical data. We stress that this 2.5 fusion-fission power-law is not the same as the critical distribution that appears at a single instant in time during a process of pure fusion but no fission. The SM gives all our statistical analyses.

The online anti-X communities often link to each other within and across platforms because of their shared interests (see SM Sec. 5 for examples). Hence clusters of interlinked anti-X communities form over time, and these clusters may also get broken up by moderators when noticed. This gives rise to additional fusion and total fission on a higher scale, where each aggregating object is now an in-built community rather than an individual, and each aggregate is a cluster-of-communities (Fig. 1(c)) as opposed to a community-of-individuals. The interaction kernel from the empirical data is again approximately product-like (Fig. 2(f) inset). Hence Eq. 1 and 2 can again be applied at this higher scale. Figure 2(e) shows that the growth of the entire movement of anti-X communities on a given platform is consistent with the theoretical shockwave solution at this higher scale.
Also, the theory's predicted 2.5-exponent power-law distribution for sizes of these clusters-of-communities is consistent with the data (Fig. 2(f)) at this higher scale. 

\begin{figure}
\centering
\includegraphics[width=0.9\linewidth]{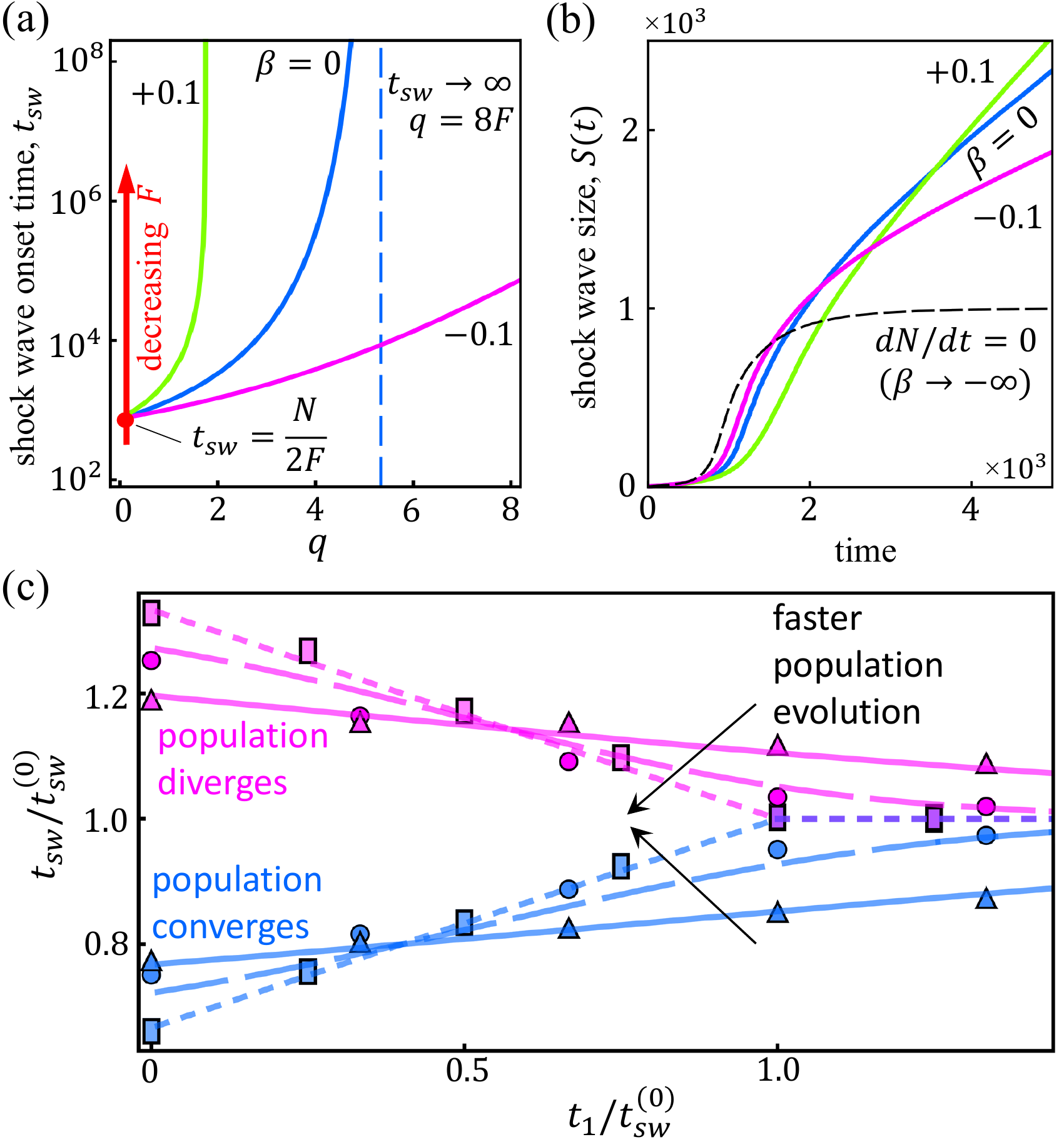}
\caption{\small{Mitigations suggested by Eq. 2 theory. (a) and (b) changing newcomer flux $\dot{N}(t)=qt^\beta$.  $\overline{F(t)}\equiv F$; $N(0)=10^3$; $q=0.5$; uniform user distribution. (c) changing user population distribution $\mathcal{P}(\{\vec{y}_{i}\})$. Starting with uniform user distribution, {\em converge} means $\mathcal{P}(\{\vec{y}_{i}\})$ evolves to single delta function (i.e. users all identical); {\em diverge} means $\mathcal{P}(\{\vec{y}_{i}\})$ evolves to two opposite peaks (i.e. users polarized). $t_1$ is the median time for the change. Analytic results (curves) and microscopic simulations (symbols).}}
\label{fig3}
\end{figure}

Equation 2 predicts how the onset time $t_{sw}$ and growth $S(t)$ can be manipulated to mitigate this anti-X behavior, as we now illustrate with two examples: 

\noindent {\em Mitigation example 1}: Change ${\dot N}(t)$, which changes the flux of new users.  Consider $\dot{N}(t)=q$. Equation 2 yields
\begin{equation}
t_{sw}=q^{-1}N(0)\left[(\alpha-1)^{2/\alpha}(\alpha+1)^{-2/\alpha}-1\right]
\label{eq:tsw-dilute}
\end{equation}
where $\alpha^2=1-8\overline{F(t)}/q$ and we assume $F(t)$ has small fluctuations compared to the mean $\overline{F(t)}$. Hence $t_{sw}$ increases with larger rate $q$: i.e. flooding the system with more heterogeneous individuals will slow the ability of a shock wave to organize macroscopically. Eventually $t_{sw}\rightarrow\infty$ for $q\geq 8\overline{F(t)}$, i.e. {\em the shockwave is prevented from forming}. 
Figure \ref{fig3}(a)(b) shows a more general example $\dot{N}(t)=qt^{\beta}$. Increasing $\beta$ with $\beta>0$, makes $t_{sw}\rightarrow \infty$ at smaller $q$. By contrast, $\beta<0$ appears to remove this transition. Figure \ref{fig3}(b) shows the corresponding growth $S(t)$. For $\beta>0$, $S(t)$ initially rises more slowly than for $\beta\leq0$, but it eventually overtakes. As $\beta$ becomes more negative, $S(t)$ rises quicker but flattens faster. 

\noindent {\em Mitigation example 2}: Change $\mathcal{P}(\{\vec{y}_{i}\})$, which changes the collective chemistry of the platform's user base and hence $F(t)$. Figure 3(c) shows two opposite cases starting from uniform $\mathcal{P}(\{\vec{y}_{i}\})$: {\em converge} to a single delta function (i.e. all identical); or {\em diverge} to two maximally separated delta functions (i.e. polarized). Both have the same simple time-dependence involving $\sigma(r,t_1,t)=(1+\exp{\left[-r(t-t_1)\right]})^{-1}$, where $r$ quantifies the rate of change and $t_1$ is the median time for the change. Equation 2 yields the approximate expression (exact for $r\rightarrow\infty$)
$t_{sw}^{{\rm converge}}\approx\frac{2}{3}t_{sw}^{(0)}+\frac{1}{3}t_{1}$ and $t_{sw}^{{\rm diverge}}\approx\frac{4}{3}t_{sw}^{(0)}-\frac{1}{3}t_{1}$, where $t_{sw}^{(0)}$ is the onset time for a static uniform $\mathcal{P}(\{\vec{y}_{i}\})$, which agrees with microscopic simulations (SM Sec. 3). 
There is also some recent empirical support: experiments show that communities formed by random aggregation from a diverse pool (i.e. uniform $\mathcal{P}(\{\vec{y}_{i}\})$ hence $F(0)=1$) are quicker to attain a high level of coherence (i.e. smaller $t_{sw}$) than those chosen to have $F(0)<1$ \cite{soc_exp19}.
The impact on $S(t)$ can be seen from an approximate solution to Eq. 2 (SM Sec. 2.2.3):
\begin{eqnarray}
\label{eq:approach-St}
S(t)&=&\frac{N(0)^{\frac{1-\alpha}{2}}N(t)^{\frac{1+\alpha}{2}}}{2\alpha}\left[\alpha-(1+2x_0)\right]+\\
& &\frac{N(0)^{\frac{1+\alpha}{2}}N(t)^{\frac{1-\alpha}{2}}}{2\alpha}\left[\alpha+(1+2x_0)\right]\nonumber\\
&-&\left[\frac{N(t)}{N(0)}\right]^{\frac{1-\alpha}{2}}\frac{z_0}{2\alpha}(\alpha+1)-\left[\frac{N(t)}{N(0)}\right]^{\frac{1+\alpha}{2}}\frac{z_0}{2\alpha}(\alpha-1)\nonumber
\end{eqnarray}
with $x_0=\omega-W(\omega e^{\omega})$, 
$z_0=\frac{N(0)}{\omega}W(\omega e^{\omega})$, and
\begin{equation}
\omega=\left[\frac{4\overline{F(t)}}{q}\right]\left[\frac{N(0)^\alpha-N(t)^\alpha}{N(t)^\alpha(\alpha+1)+N(0)^\alpha(\alpha-1)}\right]\ .
\nonumber\end{equation}

\noindent Figure 4(a) shows Eq. 4's accuracy and confirms that $S(t)$ grows slower with a later onset as $\overline{F(t)}$ decreases, i.e. making the online population more diverse will delay the anti-X shockwave onset and flatten its growth.

This theory is easily extended to include other online mechanisms. Here we summarize some of them: (1) Introducing an exponential decay $ e^{-a(s_1+s_2)}$ to the product kernel in Eq. 1 adds a non-local term $u(x+a,t)$ to Eq. 2. (SM Sec. 2.6.3) (2) Adding a multi-community ($\gamma>2$) product kernel to Eq. 1 to mimic a coordinated campaign, adds $[u(x,t)]^{\gamma-1}u'(x,t)$ to Eq. 2 (SM Sec. 2.5). (3) Individuals' own loss of interest can be mimicked by adding a monomer fragmentation term (SM Sec. 2.4.1). (4) Shifts in background population mood and influxes of new recruits in response to external events, can be mimicked by changes in $F(t)$ and $N(t)$. (5) Changes in moderator effort and/or platform (or government) tolerance can be mimicked by changing $\nu_{\rm f}$. (6) The presence of influencers can be mimicked by making the aggregation favor their particular  $\vec{y}_{i}(t)$ values. (7) The influence of external actors can be mimicked by biasing subsets of $\vec{y}_{i}(t)$ over time, as can the influence of so-called digital inoculation schemes. (8) The fact that different platforms feature different severities of harm (e.g. Facebook vs. 4Chan) and these may change in time, can be incorporated via the characters of that platform’s communities and hence its members (SM Sec. 4.2). (9) Much research has also focused on the spreading of harmful content including mis(dis)information: this could be mimicked by adding a viral process (e.g. SIR \cite{NewmanBook}) into the generative equation. However, until the correct viral process is established for online material, it makes more sense to focus on the total number of links available since this is ultimately what amplifies the traffic and encourages further growth of anti-X communities: adding $s^2 n_s(t)$ to Eq. 1 (which for large $s$ is the total number of possible links within all aggregates of size $s$) adds a diffusive term $u''(x,t)$ to Eq. 2. (10) Communities' may become coupled when their interests converge (e.g. Fig. 4(b) around the U.S. Capitol attack). Crude analytic expressions for $S(t)$ (SM Sec. 2.7) yield $N_S$ coupled differential equations for shockwaves $S_1(t)$, $S_2(t)$,..., $S_{N_S}(t)$, whose solutions agree well with the empirical data (Fig. 4(b)).

\begin{figure}
\centering
\includegraphics[width=1.0\linewidth]{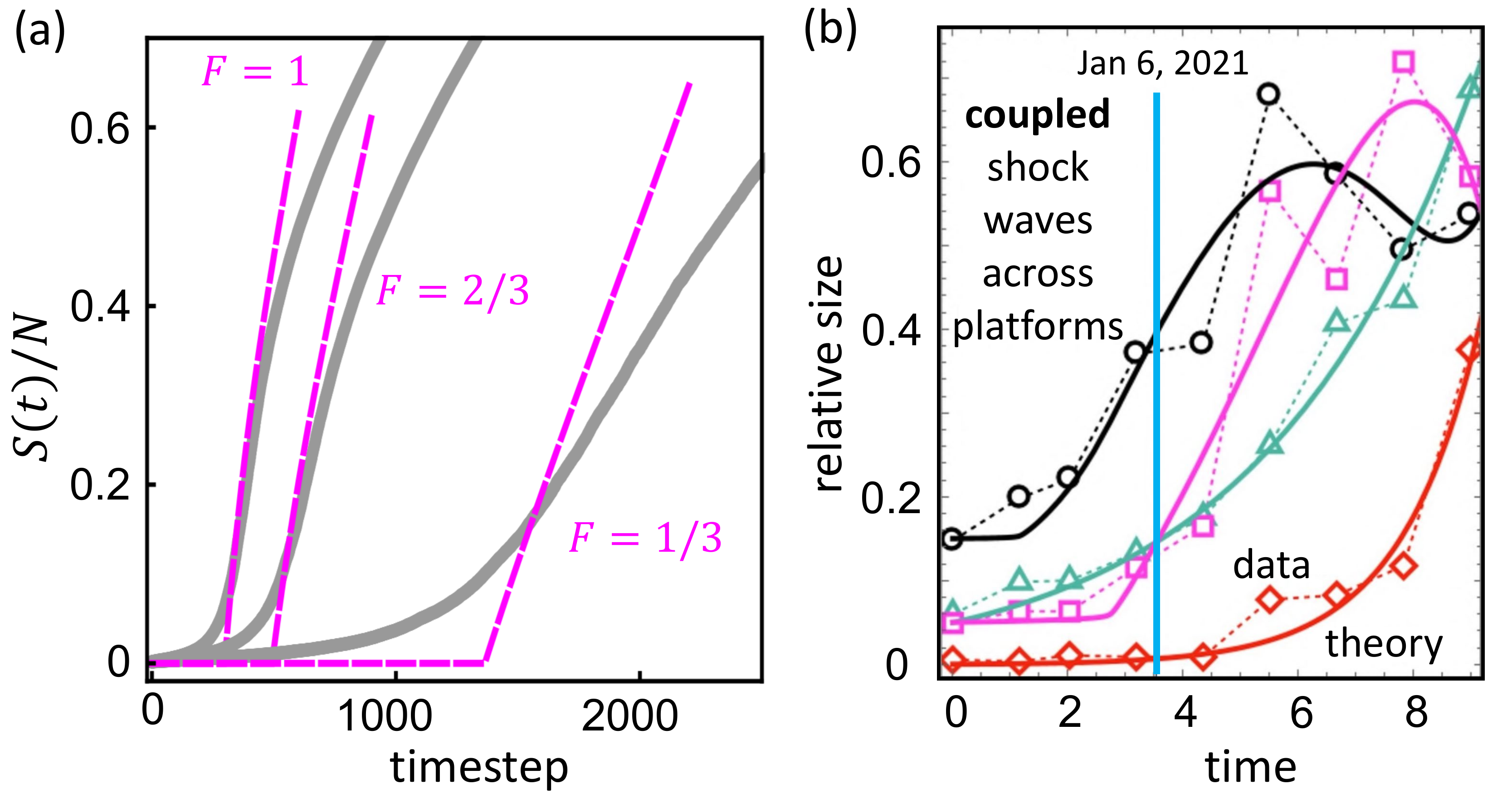}
\caption{\small{(a) Approximate analytic solutions of Eq. 2 using a time-average $\overline{F(t)}\equiv F$ (see Eq. 4, dashed magenta curves) versus exact microscopic simulations (solid gray curves show average over 100 runs). $N(0)=500$; uniform heterogeneity distribution; $\dot{N}(t)=q=0.5$. (b) Empirical data (symbols) and approximate analytic predictions (solid curves) for  coupled shock waves (SM Sec. 2.7).}}
\label{fig4}
\end{figure}

In summary, we presented a theory for online anti-X behavior. It establishes a formal connection to nonlinear fluid dynamics (e.g. shockwaves, turbulence) and hence opens a new door for physics. More empirical work is needed beyond the  blanket `anti-X' label, to understand which communities more closely follow Eq. 2. References \cite{Rick1,Rick2} show explicitly how dynamical machine-learning can help with this task, by inferring and quantifying the time-evolution of each anti-X community's `flavor' (i.e. collective traits) from its content. More broadly, our theory should in principle be applicable to many decentralized systems of heterogeneous objects. 

This research is supported by U.S. Air Force Office of Scientific Research awards FA9550-20-1-0382 and FA9550-20-1-0383.
\bibliography{SWdraft}
\end{document}